\documentclass[11pt,halfparskip]{scrartcl}

\usepackage{url}

\usepackage{mathpazo}

\usepackage{graphicx}
\usepackage[utf8]{inputenc} 
\usepackage[square]{natbib}
\usepackage{hyperref}
\usepackage{xspace}
\usepackage{booktabs}
\usepackage{fancyvrb}
\usepackage{siunitx}
\usepackage{verbatim}
\usepackage{ltablex}

\usepackage{listings}
\usepackage{color}
\definecolor{codegreen}{rgb}{0,0.6,0}
\definecolor{codegray}{rgb}{0.5,0.5,0.5}
\definecolor{codepurple}{rgb}{0.58,0,0.82}
\definecolor{backcolour}{rgb}{0.95,0.95,0.92}
\lstdefinestyle{mystyle}{
    backgroundcolor=\color{backcolour},   
    commentstyle=\color{codegreen},
    keywordstyle=\color{magenta},
    numberstyle=\tiny\color{codegray},
    stringstyle=\color{codepurple},
    basicstyle=\scriptsize,
    breakatwhitespace=false,         
    breaklines=true,                 
    captionpos=b,                    
    keepspaces=true,                 
    numbers=left,                    
    numbersep=5pt,                  
    showspaces=false,                
    showstringspaces=false,
    showtabs=false,                  
    tabsize=2
}
\newcommand{\ie}{ie}
\newcommand{\eg}{eg}
\begin{document}


%
\title{cjdb: a simple, fast, and lean database solution for the CityGML data model}

\author{
  Leon Powałka \and Chris Poon \and Yitong Xia \and Siebren Meines \and Lan Yan \and Yuduan Cai \and Gina Stavropoulou \and Balázs Dukai \and Hugo Ledoux
}
%

\date{}

\maketitle

\begin{abstract}
When it comes to storing 3D city models in a database, the implementation of the CityGML data model can be quite demanding and often results in complicated schemas.
As an example, 3DCityDB, a widely used solution, depends on a schema having 66 tables, mapping closely the CityGML architecture. 
In this paper, we propose an alternative (called `cjdb') for storing CityGML models efficiently in PostgreSQL with a much simpler table structure and data model design (only 3 tables are necessary).
This is achieved by storing the attributes and geometries of the objects directly in JSON\@.
In the case of the geometries we thus adopt the \emph{Simple Feature} paradigm and we use the structure of CityJSON\@.
We compare our solution against 3DCityDB with large real-world 3D city models, and we find that cjdb has significantly lower demands in storage space (around a factor of 10), allows for faster import/export of data, and has a comparable data retrieval speed with some queries being faster and some slower.
The accompanying software (importer and exporter) is available at \url{https://github.com/cityjson/cjdb/} under a permissive open-source license.

\end{abstract}

\section{Introduction}

The international standard \emph{City Geography Markup Language} (CityGML) is a data model designed for the storage of digital 3D models representing urban areas and landscapes \citep{Kutzner20,Groger12,OGC-CityGML3}. 
It allows us to define and store the majority of commonplace 3D objects within cities, such as buildings, roads, rivers, bridges, vegetation, and city furniture. 
Additionally, it supports various levels of detail (LoDs) for the 3D objects, which enables and facilitates complex applications and use-cases~\citep{15_ijgi_3dapps}.

The CityGML data model, currently at version 3.0, has three known encodings (more details in Section~\ref{sec:rw}):
\begin{description}
  \item[XML/GML encoding]: the XML/GML encoding (built upon GML~\citep{OGC-GML}) was initially the only standardised encoding for CityGML, which explains the---rather confusing---name choice for the data model. The latest official release of the XML/GML encoding supports CityGML version 2.0~\citep{OGC-CityGML2}; however a release planned for 2023 will also support CityGML v3.0.
  \item[CityJSON]: Its version 1.0 is standardised by the OGC~\citep{OGC-CityJSON-v10}, and its version 1.1\footnote{\url{https://cityjson.org}} implements a subset of the CityGML v3.0 data model. Its flat hierarchy and simple structure make it around 6 times more compact than the XML/GML encoding, thus allowing for easier manipulation and exchange on the web~\citep{19_ogdss_cityjson}.
  \item[3DCityDB]: the \emph{3D City Database} is a ``geo-database solution'' (schema and accompanying software) supporting three different relational DBMSs (data\-base management systems). It implements a mapping of the CityGML data model (currently for v1.0 and v2.0 only) to the database schema to allow for a fast implementation~\citep{Yao18}. It is not standardised by the OGC\@.
\end{description}

DBMSs can greatly simplify the management of large 3D city models: they are arguably the best tool to store and manage very large datasets (of any kind), are already part of the ecosystem of many organisations, and offer several advantages over file-based systems, \eg\ security, versioning, scalability, etc.~\citep{Ramakrishnan02}.
This makes 3DCityDB a popular solution, especially for handling country-level data and for offering access to multiple users. 
In most cases, the data owners store the data with 3DCityDB on a remote server and allow the users to access the data through a website, filter it by objects/LoDs/areas and obtain a subset of the 3D city model, in various different formats, for instance KML, COLLADA, and glTF\@.

However, while the 3DCityDB is widely used, \citep{Pantelios22} argues that its use can be somewhat complex and difficult for end-users.
The main culprit is the fact that datasets are split over 66 tables and the \emph{Simple Feature} paradigm~\citet{OGC-SF} is not used (geometries are stored across different tables, not in a column of an object), which translates to very complex queries that necessitate several joins.
\citet{Pantelios22} solution is to create extra views for attributes and geometries, and to offer simplified access to them through a graphical interface (built upon QGIS).
However, this comes at the cost of increasing the size of the database.

We present in this paper an alternative to 3DCityDB, which we name `cjdb'.
It is composed of a database schema (containing only 3 tables) and accompanying software for import and export of CityJSON v1.1 files (thus the CityGML v3.0 core model is supported).
As further explained in Section~\ref{sec:dm}, our data model is inspired by the \emph{Simple Feature} paradigm (each row has the geometries of the object stored in one column), but instead of using PostGIS geometry types, we exploit the fact that PosgresSQL can store JSON objects directly in binary format with the \texttt{jsonb} type.
The reason for this choice is that PostGIS geometry types (notice that to use 3D types the SFCGAL extension~\citep{SFCGAL} would be required), would not allow the storage of appearances (textures and/or materials) and of semantic information on the surfaces.
Therefore, the geometries of a given city object (\eg\ a building, a tree, a lamppost) are stored together with the object in JSON format, as defined by CityJSON\@.
Our simple structure allows us to compress by an order of around 10 the typical size of a database as stored with 3DCityDB (taking into account data and (spatial) indexes), and, as shown in Section~\ref{sec:benchmark}, this size reduction does not come with a penalty for the speed of the data retrieval. 
Our data model is at the moment only for PostgreSQL, but because it is so simple (only 3 tables are necessary), it could surely be ported to other databases.

%
\section{Related work}%
\label{sec:rw}

\subsection{CityGML data model}%
\label{sec:citygml}

To represent a region in 3D, CityGML recursively decomposes it into semantic objects~\citep{Groger12}.
It defines the classes most commonly found in an urban or a regional context, and the hierarchical relationships between them (\eg\ a building is composed of parts, which are formed of walls, which have windows). 
Also, the CityGML semantic classes are structured into several modules, \eg\ Building, Land Use, Water Bodies, and Transportation.

The geometry of the objects is realised with a subset of the geometry definitions in ISO19107~\citep{ISO19107} (only linear and planar primitives are however allowed), which also allows aggregations of geometries: a single building can for instance be modelled with a \texttt{CompositeSolid}.
Furthermore, it is possible to attach textures, materials, and semantics to each of the surfaces of a 3D geometry.
The geometry types of most geo-DBMSs do not allow us to represent such complex 3D geometries.

One of the main characteristics of CityGML is that it supports different levels of detail (LoDs) for each of the classes, which means that in theory for a single building, or a single tree, several geometries could be stored.

\subsection{CityJSON + CityJSONFeature}%
\label{sec:cityjson}

CityJSON, with its latest version 1.1, is a JSON-based exchange format for the CityGML data model.
It implements all the core modules of CityGML v3.0, and some other modules are supported\footnote{see details at: \url{https://www.cityjson.org/citygml/v30/}}.
As explained in \citet{19_ogdss_cityjson}, it was designed to improve the weaknesses of the XML-encoding of CityGML: large filesize, complex structure to manipulate, several ways to store a given characteristic, unfit for the web, etc.

Its geometry structure is similar to that of computer graphics formats (\eg\ OBJ and STL), and allows us to compress by a factor of around 6 XML-encoded CityGML files.
A thorough comparison shows that it is nearly as compact as formats that do not allow semantics, complex attributes, and coordinate reference systems~\citep{Praschl22}.

While the original files for CityJSON v1.0 were compact, one weakness was that files for large areas were not suitable for streaming. 
That is, to be able to process one object in a file, the client had to have in memory the whole file.
Version 1.1 solved this issue by introducing a new type: \texttt{CityJSONFeature}, which represents \emph{independently} one city object in a CityJSON file (\eg\ a `Building' or a `Bridge').
The idea is to decompose a region into its many features, create several JSON objects of type \texttt{CityJSONFeature}, and stream them sequentially or store them in a JSON text sequence~\citep{IETF-json-sequence}\footnote{\emph{JSON Lines text file} is one possibility: \url{https://jsonlines.org/}}.
This is conceptually the same as the \emph{GeoJSON Text Sequences}\footnote{\url{https://datatracker.ietf.org/doc/html/rfc7946\#appendix-C}} used for processing and exchanging large 2D GIS datasets.

We exploit this type to store independently each feature in one row of the database, although, as explained below, we modify the JSON structure slightly, split it over a few columns, and use indexes to accelerate performance.

\subsection{NoSQL databases}%
\label{sec:nosql}

\citet{Nys21a} developed a non-relational data model for CityJSON, and implement it in a JSON document database (MongoDB).
They tested with one dataset (containing around 3500 city objects), and managed to reduce the size by a factor of 40\% when compared to 3DCityDB\@.
Only one query was benchmarked (retrieval of a random building with its attributes and single geometry), and their solution performed better than 3DCityDB (again around 40\% faster).
There is no information about how their solution performs with other queries that practitioners expect from a DBMS solution (\eg\ the ones from Section~\ref{sec:queries}).

\subsection{3DCityDB}%
\label{sec:3dcitydb}

While the data model of CityGML could have been automatically mapped to relational tables, \citet{Yao18} mentions that for 3DCityDB a semi-automatic method was used to reduced the number of tables and the number of joins to perform queries (which will also typically reduce query times).
The result is nonetheless, for v4.4, a total of 66 tables, many of which remain empty if, for instance, only buildings without appearances are stored.

As \citet{Pantelios22} mentions, the attributes for a given type are stored in different tables, depending on whether an attribute is prescribed by the CityGML data model or not.
This complicates greatly the retrieval of information with SQL queries.

Interestingly, the 3D geometries are decomposed into their (semantic) surfaces, and each surface is stored in a separately row in a single table one table.
Different flags are used to indicate whether a 3D geometry is a solid/watertight or a surface/not.
While this approach allows to compactly store 3D volumetric geometries, in practice several joins are necessary to retrieve all the surfaces of a given feature.
The volumetric 3D types available in PostGIS-SFCGAL~\citep{SFCGAL} are also used (thus duplication of data).

Interestingly, PostGIS geometries for the footprint are not stored, only the 2D bounding box.
This is in our opinion an odd choice because many queries to 3D city models are 2D queries (all buildings inside a given area, within a given distance, etc).


%
\section{Data model, software, and engineering decisions}%
\label{sec:dm}

\subsection{Data model}

\begin{figure}[tb]
  \centering
  \includegraphics[width=0.9\linewidth]{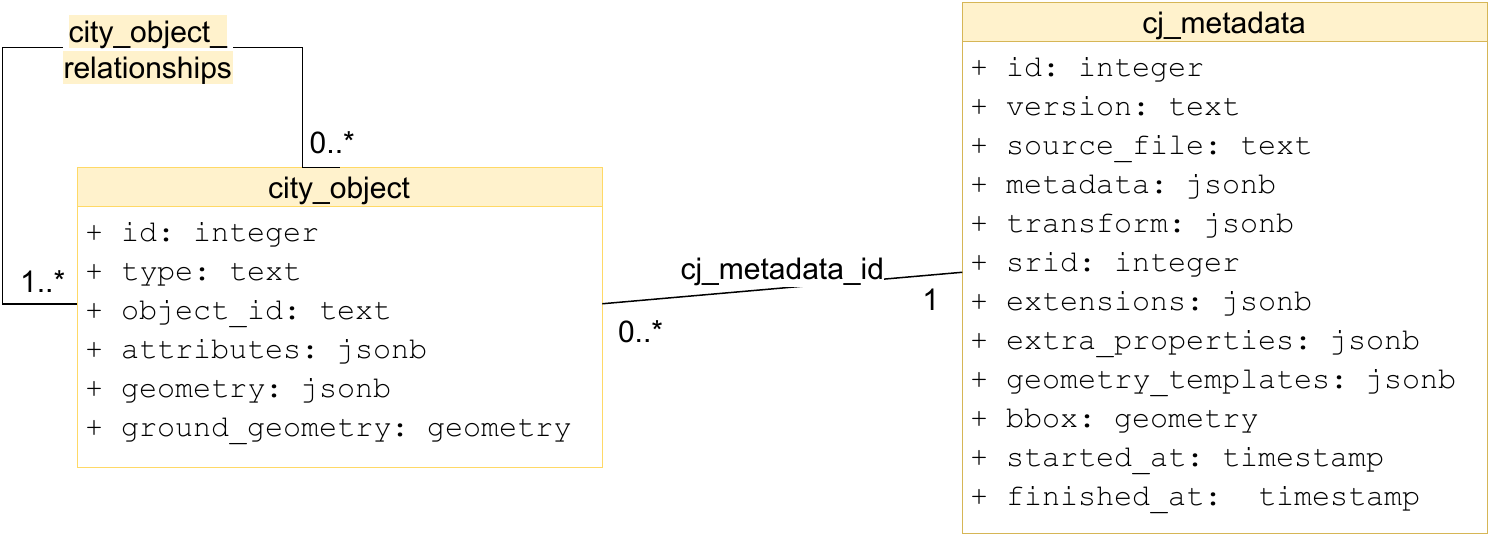}
  \caption{UML diagram of cjdb.}%
  \label{fig:uml}
\end{figure}

As shown in Figure~\ref{fig:uml}, 
the cjdb data model is simple and akin to using the Simple Feature paradigm~\citep{OGC-SF}, as PostGIS does.
Each row in the table \texttt{city\_object} stores a CityJSON city object (for instance a `Building', a `BuildingPart', a `SolitaryVegetationObject', etc.) and the `geometry' column stores a JSON array of geometries (Figure~\ref{fig:geom_column} shows an example).
\begin{figure}[tb]
  \begin{lstlisting}
   [
     {
        "type": "Solid",
        "lod": "2.2",
        "boundaries": [
           [ [[ [11.1, 22.6, 9.9], [16.21, 42.8, 19.9], ... ]
        ],
        "semantics": {
          "surfaces" : [
            { "type": "RoofSurface" }, 
            { "type": "WallSurface" },
            ...
          ],
          "values": [ [0, 1, ...] ]
        }
      }
    ]  
  \end{lstlisting}
  \caption{Example snippet stored in the `geometry' column: an array of CityJSON geometries.}%
  \label{fig:geom_column}
\end{figure}
The array is necessary because a given feature can have more than one geometry (\eg\ the most obvious case is when several LoDs are stored), and it should be noticed that the JSON stored in the database is different from CityJSON: vertices are not stored separately and we replace the vertex identifiers in the \texttt{"boundaries"} property by their real-world coordinates.

For each row, we also store the identifier of the object, its type, the attributes (stored as \texttt{jsonb}), and we also store the ground geometry as a PostGIS 2D type (to be able to spatially index them in 2D, see below).
Notice that the children objects are stored in separate rows (and not with their parent object), \ie\ a `Building' does not have its potential children in the same row, and each of them (\eg\ a `BuildingPart') is stored in a separate row.

The implemented version adds one more table: a table named \texttt{city\-\_object\-\_relation\-ships} to store the relations between `parents' and `children' city objects (CityJSON has a flat structure and stores for instance a `Building' at the same level as its `BuildingPart' or `BuildingInstallation' and links them). 
This table is added to improve query speed since we often need to process a parent with its children.

Finally, the table \texttt{cj\_metadata} has one entry for each CityJSON file imported to the database and stores the following properties of CityJSON:
\begin{itemize}
  \item the coordinate reference system (CRS);
  \item the precision used for storing the coordinates, used mostly when exporting data;
  \item the geometry templates (used for trees or bus stops);
  \item the CityJSON Extensions (to extend the base data model of CityJSON with appli\_cation-specific types/attributes/semantics);
  \item the bounding box of the file.
\end{itemize}

\subsection{Importer}%
\label{sec:importer}

To facilitate the usage of the cjdb data model we have implemented an importer and released it open-source under the MIT license: \url{https://github.com/cityjson/cjdb}\footnote{the version described in this article is v1.3}.
The tool is developed in Python and has a command-line interface.
Observe that because the data model is feature-centric, the importer will read JSONL files (JSON Sequences) and not CityJSON files.
A CityJSON file can however be automatically converted into a list of features with the accompanying software \emph{cjio}\footnote{\url{https://github.com/cityjson/cjio}}.

The importer creates the 3 necessary tables, and populates them by parsing and modifying the CityJSON features according to the cjdb data model, as explained above.

\paragraph{Ground geometry extraction:}
As many queries on a 3D city model are typically performed in 2D, such as retrieving all objects within a certain area or selecting the object clicked upon in a 2D view, we have chosen to store the ground surface of each object as a 2D PostGIS geometry.
This is achieved by iterating over all of the object's surfaces and selecting the horizontal ones with the lowest elevation.
If multiple levels of detail (LoDs) are available, we select the lowest LoD.
This addition enables us to perform rapid 2D spatial queries on the data without any joins (see Section~\ref{sec:benchmark}).
In comparison, performing the same spatial queries in 3DcityDB requires multiple joins.
Alternatively, the enveloping bounding box, which in 3DCityDB is stored together with the object, can be used instead of the actual object geometry in order to perform spatial queries when the accuracy is not important.

\paragraph{Indexes:}
The data in the \texttt{city\_object} table is expected to be retrieved mostly through spatial queries.
Therefore, we decided to add a GiST index on the `ground\_geometry' column and cluster the table based on that index.
In order to improve query performance on the JSON columns we added a Generalized Inverted index (GIN), which is specialised for items with composite values.
Additional full or partial BTree indexes can be applied during import, on specific attributes of the city objects, if the user expects that the table will be queried based on those attributes.

\subsection{Exporter}

The Python implementation of cjdb also offers an exporter.
As is the case for 3DCityDB, a SQL query is used to filter which objects in the database should be exported (the identifiers in the table \texttt{city\_object}).
The output is a CityJSONL file (a sequence of \texttt{CityJSONFeatures}), which can automatically be converted to a CityJSON file with cjio.

%
\section{Benchmark}%
\label{sec:benchmark}

To compare the performance of the cjdb data model against that of 3DCityDB, we created a benchmark dataset using data from 3 different countries; the Netherlands (3DBAG), Austria (Vienna), and USA (NYC).
The 3DBAG dataset is composed of 100 tiles from the 3DBAG\footnote{\url{www.3dbag.nl}}~\citep{Peters22}; we randomly chose tiles from 3 cities in the Netherlands (Delft, Amersfoort, and Zwolle).
The Vienna dataset covers the Austrian city whereas the NYC dataset covers a small part of central New York City; both datasets can be downloaded in CityJSON format from \url{https://www.cityjson.org/datasets/}.
As can be seen in Table~\ref{tab:dt_stats}, all the datasets are building-centric, as is often the case with 3D city models, but here they are modelled differently and have different LoDs/sizes. 
\begin{table}
  \centering
  \caption{The 3 datasets used for the benchmark.}
  \small
  \begin{tabular}
    {@{}lrrcr@{}}\toprule
     & {\# \texttt{Building}} &  {\# \texttt{BuildingPart}}  & LoDs present & \# attributes  \\
    \midrule
    \textbf{3DBAG}          & \num{112673} & \num{110387}  & 0/1.2/1.3/2.2  & 30  \\
    \textbf{NYC}            & \num{23777}  & \num{0}       & 2              & 3   \\
    \textbf{Vienna}         & \num{307}    & \num{1015}    & 2              & 7   \\
    \bottomrule
  \end{tabular}%
  \label{tab:dt_stats}
\end{table}

We imported each dataset in two different databases, one created with 3DCityDB and one cjdb, and below we compare them in terms of import/export time, data size, and data retrieval.
Since cjdb is only available for PostgreSQL, we did not perform any tests on Oracle or PolarDB\@.

\subsection{Import and export times}

\begin{table}
  \centering
  \caption{Import and export times, from/to CityJSONL\@. All times in seconds.} 
  \small
  \begin{tabular}{@{}lrrrrrr@{}} \toprule
    && \multicolumn{2}{c}{\textbf{3DCityDB}} && \multicolumn{2}{c}{\textbf{cjdb}} \\
    \cmidrule{3-4} \cmidrule{6-7} 
    && import & export && import & export \\
    \midrule
    \textbf{3DBAG}  && \num{6780} & \num{721} && \num{1260} & \num{412} \\
    \textbf{NYC}    &&  \num{273} & \num{161} &&   \num{23} &  \num{25}  \\
    \textbf{Vienna} &&   \num{12} &   \num{7} &&    \num{2} &   \num{2.5}   \\
    \bottomrule
  \end{tabular}%
  \label{tab:io}
\end{table}

We compared the import time for all 3 datasets and we found that cjdb is considerably faster than 3DCityDB\@.
As an example, the 100 tiles of the 3DBAG were imported in \qty{21}{\minute} in cjdb whereas it took \qty{113}{\minute} with 3DCityDB\@; see Table~\ref{tab:io} for all details.
This is expected, since the storage in the cjdb database is very close to the data model of CityJSONL, while for 3DCityDB the geometries need to be processed and each surface to be stored separately.
However, it is worth noting that in 3DCityBD the creation of the necessary tables is performed separately before the import, whereas cjdb creates the necessary tables at the time of import. 

For the export, we exported each dataset to a CityJSON file (for 3DCityDB) and to a CityJSONL file (for cjdb).
As shown in Table~\ref{tab:io}, cjdb is also generally faster.

Observe that those values are somewhat unfair to compare because the output is to a different format (and some time would be needed to convert from a CityJSONL to a CityJSON), because different languages are used for the importer/exporter (Java for 3DCityDB, Python for cjdb), and because 3DCityDB exporter is multi-threaded.

\subsection{Database size}

One significant difference between 3DCityDB and cjdb is the database size they occupy. 
After importing the datasets, we measured the total size of each database, including the indexes and the TOAST tables (The Oversized-Attribute Storage Technique), as shown in Table~\ref{tab:size_comparison}.
We notice that cjdb occupies significantly less space that what 3DCityDB demands for the same data.

The main reason for this size difference lies in the different approaches to storing the features.
In 3DCityDB, the different semantic surfaces (wall/roof/ground/etc) are considered separate city objects, contributing to the total number of rows in the \texttt{cityobject} table.
In cjdb, the walls, roof and surfaces are considered intrinsic properties of each city object and thus remain in the \texttt{jsonb} format stored at the geometry column of each object.
This decision has several advantages: reduction of the size of the \texttt{city\_objects} table and faster and simpler queries where only city objects are concerned (\eg\ Q1 and Q4 in Table~\ref{tab:allqueries}).
But there is also a major drawback: we cannot perform spatial queries on the geometries of specific semantic surfaces.

Another reason for the different database sizes is the different amount of indexes.
We notice that cjdb implements significantly less indexes and as a result requires way less storage for them.
In 3DCityDB, indexes account for almost half of the database's size.  

One more noticeable difference is that, since cjdb stores attributes and geometries in \texttt{jsonb} format, it requires more space for TOAST tables.
TOAST (\emph{The Oversize Attribute Storage Technique}) is a PostgreSQL mechanism which controls the size of the data stored in a field.
If the data exceeds the maximum allowed limit, TOAST breaks the too-wide field values down into smaller pieces, and stores them out-of-line in a TOAST table.
Columns of type \texttt{jsonb} tend to carry quite wide values and often utilise the TOAST tables.
Thus when measuring the cjdb database size we need to take the extra toast tables into account. 
However, even with the extra TOAST tables, the total database size remains around ten times smaller than that of 3DCityDB\@. 

Finally, it should be noticed that the 3DCityDB size should in reality be larger than the number we obtain: since its data model allows only but one geometry per LoD and that the refined LoDs by \citet{Biljecki16c} are not supported, the 3DCityDB importer selects either the LoD1.2 or LoD1.3 from the inputs, and thus one LoD is missing.
In cjdb, all available LoDs are stored together in the `geometry' column.

\begin{table}
  \centering
  \caption{Database size comparison for 100 tiles of the 3DBAG dataset, all values in MB.} 
  \small
  \begin{tabular}{@{}lrrrrrrrrrr@{}} \toprule
    && \multicolumn{4}{c}{\textbf{3DCityDB}} && \multicolumn{4}{c}{\textbf{cjdb}} \\
    \cmidrule{3-6} \cmidrule{8-11} 
    && tables & indexes & TOAST & total && tables & indexes & TOAST & total \\
    \midrule
    \textbf{3DBAG}  && 5463 & 4322 & 112 & 9898 && 257 & 57 & 755 & 1070 \\
    \textbf{NYC}    && 590 & 735 & 0.5 & 1326 && 26 & 4 & 25 & 54 \\
    \textbf{Vienna} && 30 & 42 & 0.5 & 73 && 1.5 & 0.5 & 4 & 6 \\
    \bottomrule
  \end{tabular}%
  \label{tab:size_comparison}
\end{table}

\subsection{Data Retrieval}%
\label{sec:queries}

The data retrieval comparison was performed based on the execution time of SQL queries which aim to retrieve the same data from both databases. 
Postgres heavily relies on caching, therefore the queries below were run several times to ensure the cache was warm.

We performed 8 queries that we believe are representative of what a typical practitioner (or server hosting data to be downloaded) would need.

Those are listed in Table~\ref{tab:allqueries}.
The exact SQL queries we used for the 3DBAG dataset are listed in Appendix~\ref{sec:sqlqueries}; similar queries were used for the other 2 datasets.

\begin{table}
  \centering
  \caption{The 8 queries we used for the benchmark.}%
  \begin{tabular}{@{}ll@{}}
    \toprule
     Q1 & Retrieve the ids of all buildings based on one attribute (roof height higher than \qty{20}{m}) \\
     Q2 & Retrieve all buildings within a 2D bounding box \\
     Q3 & Retrieve building intersecting with a 2D point \\
     Q4 & Retrieve the number of parts for each building \\
     Q5 & Retrieve all buildings having a specific LoD geometry \\
     Q6 & Add new `footprint\_area' attribute \\
     Q7 & Update `footprint\_area' attribute by adding 10m \\
     Q8 & Delete `footprint\_area' attribute \\
     \bottomrule
  \end{tabular}%
  \label{tab:allqueries}
\end{table}

\paragraph{Q1. Query based on attributes:} 

3DCityDB offers a list of predefined building attributes within the \texttt{building} table, which include `year\_of\_construction' and `roof\_type'---attributes that are not in this list are stored in the table \texttt{cityobject\_genericattrib}.
Cjdb on the other hand offers more flexibility since all the attributes remain in JSON format in the attributes column, regardless of the attribute name. 

Since none of our datasets have attributes from the 3DCityDB's predefined list, we decided to compare the attribute-based data retrieval for both databases based on non-listed attributes.
In this specific example, we queried all the buildings with roof height (`h\_dak\_max' for BAG `HoeheDach' for Vienna) higher than \qty{20}{m}.
The New York dataset was not taken into account for this query, since there is no specific attribute about the roof height. 

For cjdb no join is necessary since the attributes are stored together with the city object but the equivalent in 3DCityDB requires a join between the \texttt{city\_object} and the \texttt{cityobject\_genericattrib} tables.
As shown in Table~\ref{tab:exec_time}, cjdb it is faster than 3DcityDB for Vienna but performs almost the same as 3DcityDB for the 3DBAG dataset.
This is related to both the size of the dataset and the size of the attributes column in cjdb; the bigger the \texttt{jsonb} column, the slower the query, since the information stored in the TOAST tables will need to be retrieved and decompressed.

\paragraph{Q2 \& Q3. 2D Spatial queries:} 
The spatial queries in 3DCityDB tend to be complicated since the geometries of the objects are stored in other tables and require joins to retrieve them.
As an example, in order to find all buildings within a certain 2D bounding box, their ground surfaces had to be retrieved from the \texttt{surface\_geometry} table.
An alternative but less accurate solution would be to use the bounding box geometry of the object, which is stored in the city objects' table.
Cjdb does not require any join to retrieve the same data, since the \texttt{city\_objects} table contains the ground geometries of the objects.
As a result the cjdb query is significantly faster than the  equivalent of 3DCityDB, as shown in Table~\ref{tab:exec_time}.
We observe similar speed differences with other spatial queries, such as retrieving the building intersecting with a given point in 2D (Q3).

\paragraph{Q4. Number of parts query:} 
We compared how the two databases perform with the retrieval of the number of building parts per building.
For simplicity we considered only first level children for each building and we also require the buildings without any parts to be part of the result. 
Both queries require a single join and their execution times varies depending on the size of the dataset.
When the 3DBAG dataset is examined, cjdb seems to be slower than 3DCityBD but for the other datasets, cjdb is faster.
However it is worth mentioning that for cjdb the join with the city object table could be skipped and the number of parts could be retrieved with a single aggregation query on the \texttt{city\-\_object\-\_relationships} table, but only for the objects which have parts.

\paragraph{Q5. LoD-based query:} 
While one strength of CityGML is that many LoDs of one city object can be stored with the object, exporting a given one to perform an analysis is useful in practice.  
We therefore tested a query to obtain all building ids with LoD1 (in 3DCityDB) and LoD1.2 (in cjdb) for the 3DBAG dataset and LoD2 for the other 2 datasets.
Cjdb performs slower for all datasets; this is probably due to the large size of the `geometry' column which is in \texttt{jsonb} format and therefore requires joining to the TOAST tables.

It should be noticed that if the geometry of the building needs also to be retrieved, the equivalent query for 3DCityBD requires joins which significantly lower the speed. 

\paragraph{Q6, Q7 \& Q8. INSERT/UPDATE/DELETE attribute queries:} 
We also compared how the databases perform when adding, modifying or deleting an attribute of a building.
When it comes to adding new attributes to the database (Q6), cjdb performs considerably slower than 3DcityDB\@.
This can be attributed to the different structure of the databases: new attributes in 3DCityDB can simply be inserted as rows in the relevant table, whereas in cjdb the \texttt{jsonb} must be modified. 
However, when it comes to updating existing attributes (Q7), the speed of cjdb is similar for datasets having many attributes, and faster for datasets having few attibutes.
Deleting attributes (Q8) follows a similar behaviour.
Generally, it can be noticed that the number of attributes and the size of the dataset can significantly affect queries on the \texttt{jsonb} columns.

\begin{table}
  \centering
  \caption{Average execution times for the benchmark queries (all times in ms).} 
  \begin{tabular}{@{}lrrrrrrrrr@{}} \toprule
    && \multicolumn{2}{c}{\textbf{3DBAG}} && \multicolumn{2}{c}{\textbf{NYC}} && \multicolumn{2}{c}{\textbf{Vienna}} \\
    \cmidrule{3-4} \cmidrule{6-7} \cmidrule{9-10}  
    && 3DCityDB & cjdb && 3DCityDB & cjdb && 3DCityDB & cjdb \\
    \midrule
    \textbf{Q1}  &&  \num{290} & \num{285}  && -- & --  && \num{22} & \num{2} \\
    \textbf{Q2}  &&  \num{172} & \num{1.4}  && \num{59} & \num{0.3} && \num{14} & \num{0.4} \\
    \textbf{Q3}  &&  \num{1.0} & \num{0.2}  && \num{0.4} & \num{0.1} && \num{0.2} & \num{0.1} \\
    \textbf{Q4}  &&  \num{418} & \num{478}  && \num{240} & \num{46} && \num{15} & \num{2} \\
    \textbf{Q5}  &&  \num{343} & \num{1877}  && \num{217} & \num{392} && \num{15} & \num{34} \\
    \textbf{Q6}  &&  \num{3981} & \num{11660}  && \num{1135} & \num{1425} && \num{15} & \num{10} \\
    \textbf{Q7}  &&  \num{10796} & \num{10903}  && \num{2333} & \num{1161} && \num{15} & \num{9} \\
    \textbf{Q8}  &&  \num{4393} & \num{10984}  && \num{1040} & \num{682} && \num{59} & \num{8} \\
    
    \bottomrule
  \end{tabular}%
  \label{tab:exec_time}
\end{table}

\section{Conclusions and future work}%
\label{sec:discussion}

The cjdb project started with the goal of creating a simpler and leaner alternative to 3DCityDB for web servers, allowing users to efficiently store and retrieve 3D city models.
Our data model follows the Simple Feature paradigm and has only 3 tables (instead of 66 for 3DCityDB).
This is achieved by maintaining the structure of CityJSON and storing JSON directly in the database, using the PostgreSQL type \texttt{jsonb}.
Because the structure of the data model is close to that of CityJSON files, we can significantly improve the import/export times to/from a database.
Furthermore, as we have shown, our simple model is around ten times more compact and it offers retrieval speed comparable to those of 3DCityDB\@.
More specifically we notice that the cjdb performs better when it comes to 2D spatial queries but it performs slower when the \texttt{jsonb} columns need to be altered (up to 3 times slower).
We also notice that the query speed on the \texttt{jsonb} columns is significantly affected by the size of the dataset; the more the attributes in the \texttt{jsonb} columns get, more time is required to parse them.

\begin{figure}
  \centering
  \includegraphics[width=0.8\linewidth]{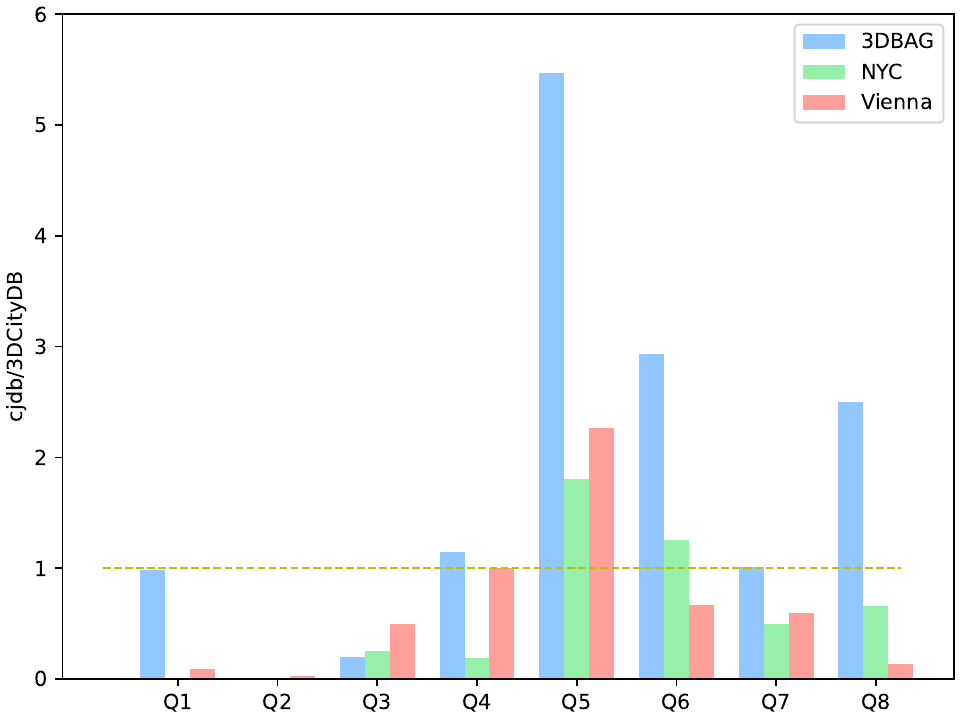}
  \caption{Data retrieval comparison between cjdb and 3DCityBD for the 8 queries using the 3DBAG dataset. The y-axis corresponds to (cjdb/3DCityDB); a bar lower than the yellow line (located at 1.0) means that cjdb has a faster query time, a bar higher than 1.0 means a slower query time.}%
  \label{fig:performance}
\end{figure}

Figure~\ref{fig:performance} shows the 8 queries, each time have been normalised (we divided the time of cjdb by that of 3DCityDB).

It should be mentioned that the data model of 3DCityDB is more generic and can be implemented with 3 different DBMSs (and not only PostgreSQL).
It also allows us to query semantic surfaces, something that is currently not possible with the cjdb model because it stores the semantic surfaces in \texttt{jsonb} format in the `geometry' column of each city object.
However, we plan to remedy to the situation by implementing database functions to extract the geometries from semantic surfaces.
Furthermore, at the moment, retrieving data from cjdb requires the user to parse the \texttt{jsonb} and extract the necessary information, something that is far from being optimal.
We plan in the near future to implement some helper plugins in mainstream open-source products (\eg\ QGIS) to be able to view and query cjdb, similar to what is currently being built for 3DCityDB\footnote{\url{https://github.com/tudelft3d/3DCityDB-Tools-for-QGIS}}.

While this was not previously discussed, it should be mentioned that CityJSON Extensions are supported by the cjdb data model.
This means that extra functions to dynamically extend the data model (as required for 3DCityDB, see \citet{Yao17}) are not necessary.
This is because CityJSON Extensions, unlike CityGML Application Domain Extensions (ADEs), are constrained to follow the structure and rules of other city objects (see \citet{19_ogdss_cityjson} for more details).

We also plan to add support for textures and material, at the moment the information is simply stored in the JSON of each geometry, but as is the case for semantic surfaces, we will add database functions to allow users to query and update those.


%
\bibliographystyle{spbasic}
\bibliography{main}

\newpage
\appendix
\section{The 8 SQL queries used for the benchmark  for the 3DBAG dataset.}%
\label{sec:sqlqueries}

\begin{table}[h]
  \centering
  \begin{tabular}{@{}p{10mm}p{55mm}p{55mm}@{}}
    \toprule
     & \textbf{3DCityDB} & \textbf{cjdb} \\
     \midrule
     Q1 & \tiny
\begin{verbatim}
SELECT
    gmlid
FROM
    3dcitydb.cityobject co
    JOIN 3dcitydb.cityobject_genericattrib coga
    ON co.id = coga.cityobject_id
WHERE
    coga.attrname = 'h_dak_max'
    AND coga.realval > 20;
\end{verbatim} & \tiny
\begin{verbatim}
SELECT
    object_id
FROM
    cjdb.city_object
WHERE
    (attributes -> 'h_dak_max') :: float > 20;
\end{verbatim} \\
     \midrule
     Q2 & \tiny
\begin{verbatim}
SELECT
    c.gmlid,
    sg.geometry
FROM
    3dcitydb.thematic_surface AS ts
    LEFT JOIN 3dcitydb.surface_geometry AS sg
    ON ts.lod2_multi_surface_id = sg.parent_id
    LEFT JOIN 3dcitydb.cityobject AS c
    ON ts.building_id = c.id
WHERE
    ST_Contains(
        ST_MakeEnvelope(
            85400,
            446900,
            85600,
            447100,
            7415
        ),
        sg.geometry
    )
    AND ts.objectclass_id = 35;
\end{verbatim} & \tiny
\begin{verbatim}
SELECT
    object_id,
    ground_geometry
FROM
    cjdb.city_object
WHERE
    TYPE = 'Building'
    AND ST_Contains(
        ST_MakeEnvelope(
            85400,
            446900,
            85600,
            447100,
            7415
        ),
        ground_geometry
    );
\end{verbatim} \\
     \midrule
     Q3 & \tiny
\begin{verbatim}
SELECT
    c.gmlid,
    sg.geometry
FROM
    3dcitydb.thematic_surface AS ts
    LEFT JOIN 3dcitydb.surface_geometry AS sg
    ON ts.lod2_multi_surface_id = sg.parent_id
    LEFT JOIN 3dcitydb.cityobject AS c
    ON ts.building_id = c.id
WHERE
    sg.geometry && st_setsrid(ST_MakePoint(85200, 446900), 7415)
    AND ts.objectclass_id = 35;
\end{verbatim} & \tiny
\begin{verbatim}
SELECT
    object_id,
    ground_geometry
FROM
    cjdb.city_object
WHERE
    "type" = 'Building'
    AND ground_geometry 
    && st_setsrid(ST_MakePoint(85200, 446900), 7415);
\end{verbatim} \\
    \bottomrule
  \end{tabular}%
  \label{tab:sql1}
\end{table}
\begin{table}
  \centering
  \begin{tabular}{@{}p{10mm}p{55mm}p{55mm}@{}}
    \toprule
     & \textbf{3DCityDB} & \textbf{cjdb} \\
     \midrule
     Q4 & \tiny
\begin{verbatim}
SELECT
    cj.gmlid AS building_id,
    count(b.id) AS number_of_parts
FROM
    3dcitydb.cityobject cj
    LEFT JOIN 3dcitydb.building b
    ON cj.id = b.building_parent_id
WHERE
    cj.objectclass_id = 26
GROUP BY
    cj.id;
\end{verbatim} & \tiny
\begin{verbatim}
SELECT
    co.object_id AS building_id,
    COUNT(cor.child_id) AS number_of_parts
FROM
    cjdb.city_object co
    LEFT JOIN cjdb.city_object_relationships cor
    ON co.object_id = cor.parent_id
WHERE
    co.type = 'Building'
GROUP BY
    co.object_id;
\end{verbatim} \\
     \midrule
     Q5 & \tiny
\begin{verbatim}
SELECT
    co.gmlid
FROM
    3dcitydb.cityobject co
    JOIN 3dcitydb.building b
    ON co.id = b.id
WHERE
    b.lod1_solid_id IS NOT NULL;
\end{verbatim} & \tiny
\begin{verbatim}
SELECT
    object_id
FROM
    cjdb.city_object
WHERE
    geometry::jsonb @> '[{"lod": 1.2}]'::jsonb;
\end{verbatim} \\
     \midrule
     Q6 & \tiny
\begin{verbatim}
INSERT INTO
    3dcitydb.cityobject_genericattrib (attrname, datatype, realval, cityobject_id)
SELECT
    'footprint_area',
    3,
    ST_Area(cityobject.envelope),
    cityobject.id
FROM
    3dcitydb.cityobject
WHERE
    cityobject.objectclass_id = 26;
\end{verbatim} & \tiny
\begin{verbatim}
UPDATE
    cjdb.city_object
SET
    attributes = jsonb_set(
        attributes :: jsonb,
        '{footprint_area}',
        to_jsonb(ST_Area(ground_geometry))
    ) :: json
WHERE
    TYPE = 'Building';
\end{verbatim} \\
     \midrule
     Q7 & \tiny
\begin{verbatim}
UPDATE
    3dcitydb.cityobject_genericattrib
SET
    realval = realval + 10
WHERE
    attrname = 'footprint_area';
\end{verbatim} & \tiny
\begin{verbatim}
UPDATE
    cjdb.city_object
SET
    attributes = jsonb_set(
        attributes :: jsonb,
        '{footprint_area}',
        to_jsonb(
            CAST (
                jsonb_path_query_first(
                    attributes :: jsonb,
                    '$.footprint_area'
                ) AS real
            ) + 10.0
        )
    ) :: json
WHERE
    TYPE = 'Building';
\end{verbatim} \\
     \midrule
     Q8 & \tiny
\begin{verbatim}
DELETE FROM
    3dcitydb.cityobject_genericattrib
WHERE
    attrname = 'footprint_area';
\end{verbatim} & \tiny
\begin{verbatim}
UPDATE
    cjdb.city_object
SET
    attributes = jsonb_set_lax(
        attributes :: jsonb,
        '{footprint_area}',
        NULL,
        TRUE,
        'delete_key'
    ) :: json
WHERE
    TYPE = 'Building';
\end{verbatim} \\
    \bottomrule
  \end{tabular}%
  \label{tab:sql2}
\end{table}

\end{document}